\documentclass[letterpaper,twocolumn,preprintnumbers,english,aps,prl,superscriptaddress]{revtex4-1}
\usepackage[T1]{fontenc}
\usepackage[latin9]{inputenc}
\usepackage{verbatim}
\usepackage{amsmath}
\usepackage{amssymb}
\usepackage{graphicx,color}
\usepackage{ulem}
\makeatletter

\pdfpageheight\paperheight
\pdfpagewidth\paperwidth

\usepackage{babel}

\makeatother

\usepackage{babel}
\begin{document}
\title{Lorentz-covariant kinetic theory for massive spin-1/2 particles}
\author{Xin-Li Sheng}
\affiliation{Key Laboratory of Quark and Lepton Physics (MOE) and Institute of
Particle Physics, Central China Normal University, Wuhan, 430079,
China}
\author{Qun Wang}
\affiliation{Interdisciplinary Center for Theoretical Study and Department of Modern
Physics, University of Science and Technology of China, Hefei, Anhui
230026, China}
\affiliation{Peng Huanwu Center for Fundamental Theory, Hefei, Anhui 230026, China}
\author{Dirk H.\ Rischke}
\affiliation{Institute for Theoretical Physics, Goethe University, Max-von-Laue-Str.\ 1,
D-60438 Frankfurt am Main, Germany}
\affiliation{Helmholtz Research Academy Hesse for FAIR, Campus Riedberg, Max-von-Laue-Str.\ 12,
D-60438 Frankfurt am Main, Germany}

\begin{abstract}
We construct a matrix-valued spin-dependent distribution function
(MVSD) for massive spin-1/2 fermions and study its properties under Lorentz transformations.
Such transformations result in a Wigner rotation in spin space and in a nontrivial
matrix-valued shift in space-time, which corresponds to the side jump in the massless case.
We express the vector and axial-vector components of the Wigner function in terms of
the MVSD and show that they transform in a Lorentz-covariant manner.
We then construct a manifestly Lorentz-covariant Boltzmann equation
which contains a nonlocal collision term encoding spin-orbit coupling.
Finally, we obtain the spin-dependent distribution function in local equilibrium by demanding
detailed balance.
\end{abstract}

\preprint{USTC-ICTS/PCFT-22-06}
\maketitle

\paragraph{Introduction. --- }

In noncentral heavy-ion collisions, a part of the large orbital
angular momentum (OAM) of the system is converted into
polarization of final-state hadrons \cite{Liang:2004ph,Liang:2004xn,Voloshin:2004ha}.
The polarization of $\Lambda$ and $\bar{\Lambda}$
hyperons along the direction of the global OAM has been experimentally measured
by the STAR collaboration \cite{STAR:2017ckg,Adam:2018ivw} and
agrees well with theoretical calculations \cite{Becattini:2013fla,Karpenko:2016jyx,Wei:2018zfb}.
However, the dependence on azimuthal angle of the
longitudinal polarization of $\Lambda$'s \cite{STAR:2019erd}
shows the opposite sign as in theoretical frameworks which reproduce the global polarization.
Various efforts \cite{Florkowski:2019qdp,Sun:2018bjl,Florkowski:2019voj,Becattini:2019ntv,Xia:2019fjf,Wu:2019eyi,Liu:2019krs}
have been made to resolve this so-called ``sign problem of the
longitudinal polarization'', but a fully convincing explanation does not yet
exist. It was recently proposed \cite{Becattini:2021suc,Liu:2021uhn,Fu:2021pok,Becattini:2021iol} that
previous calculations had missed a shear-induced contribution
to the polarization at freeze-out, which has the potential to solve
this problem, but so far results including this term appear to be sensitive to the equation
of state and other parameters of the calculation \cite{Yi:2021ryh}.

Therefore, from both the theoretical and the experimental perspective, a consistent
way to describe the dynamics of spin in heavy-ion collisions is urgently needed. Recently,
a lot of activity was devoted to deriving kinetic theory for massive particles with spin
\cite{Zhang:2019xya,Li:2019qkf,Liu:2020ymh,Yang:2020hri,Weickgenannt:2020aaf,Li:2020cwq,Manuel:2021oah,Weickgenannt:2021cuo,Sheng:2021kfc,Lin:2021mvw}
and spin hydrodynamics \cite{Florkowski:2017ruc,Florkowski:2017dyn,Montenegro:2017rbu,Florkowski:2018fap,Florkowski:2018myy,Becattini:2018duy,Hattori:2019lfp,Bhadury:2020puc,Montenegro:2020paq,Li:2020eon}.
The dynamics of massless particles is described by Chiral Kinetic
Theory (CKT), first proposed in Refs.\ \cite{Son:2012wh,Stephanov:2012ki,Chen:2012ca}.
The helicity, defined as the product of momentum and spin, is Lorentz invariant.
However, both the momentum and the spin of a particle will change under a
Lorentz boost, in order to preserve the helicity. Conservation of
total angular momentum then requires that the OAM of the particle also changes.
This in general implies that the particle's position will undergo a nontrivial shift,
which is called the side-jump effect \cite{Chen:2014cla,Chen:2015gta,Hidaka:2016yjf}.
On the other hand, it has been realized that the collisionless kinetic theory for
massive particles can be smoothly connected to CKT, if one properly
defines the reference frame \cite{Sheng:2020oqs} or if one directly replaces
the spin vector by the momentum vector \cite{Gao:2019znl,Weickgenannt:2019dks,Wang:2019moi,Hattori:2019ahi}.
Such a connection exists because Wigner's little group for massless
particles can be obtained from that for massive particles by taking
the infinite-momentum limit and the massless limit at the same time,
indicating that spin for massive particles reduces to helicity
in this limit \cite{Wigner:1939cj,Kim:1986gq,Kim:1989wt}. Then,
one naturally expects that the massless limit for
the collision term as well as for the equilibrium
distribution agrees with the result from CKT \cite{Chen:2015gta,Hidaka:2016yjf,Hidaka:2017auj}.
In order to confirm this expectation, we need to discard any reference
to the rest frame of a massive particle, as such a frame does not exist for massless
particles. Instead, we need to consider a Lorentz boost between two arbitrary reference
frames.

In this work, we derive a matrix-valued spin-dependent
distribution function (MVSD) and show that its Lorentz-transformation properties are highly
nontrivial: in addition to a Wigner rotation in spin space, the MVSD
undergoes a matrix-valued shift in space-time, which is similar
to the side-jump effect for massless particles. Using this MVSD, we then construct
a Lorentz-covariant Boltzmann equation with a nonlocal collision term, which forms
the theoretical foundation for a consistent description of the dynamics of massive
spin-1/2 particles in heavy-ion collisions.
It constitutes a well-founded theory for numerical simulations of spin polarization and thus may
potentially contribute to solving the sign problem of the longitudinal polarization.

\paragraph{Matrix-valued spin-dependent distribution function. --- }

We define a plane-wave state as
\begin{equation} \label{eq:plane_wave_state}
|\mathbf{p},s \rangle \equiv a_{\mathbf{p},s}^{\dagger} |0\rangle\;,
\end{equation}
where $a_{\mathbf{p},s}^{\dagger}$ is the creation operator for a particle with
momentum $\mathbf{p}$ and spin $s$, with the corresponding annihiliation
operator being $a_{\mathbf{p},s}$, fulfilling the anticommutation relation
$\left\{ a_{\mathbf{p},s}^{\dagger}, a_{\mathbf{p}^\prime,s^\prime} \right\}
= 2E_\mathbf{p} (2 \pi \hbar)^3 \delta^{(3)}( \mathbf{p} - \mathbf{p}^\prime) \delta_{ss'}$,
with the mass-shell energy $E_{\bf p}\equiv\sqrt{{\bf p}^{2}+m^{2}}$.
The density matrix is defined as
\begin{equation}
\rho\equiv\sum_{rs}\int Dp_{1}\int Dp_{2}\, \tilde{f}_{rs}(\mathbf{p}_{1},\mathbf{p}_{2})
\left|\mathbf{p}_{1},r\right\rangle \left\langle \mathbf{p}_{2},s\right| \;,
\label{eq:density_matrix}
\end{equation}
where the invariant momentum-integration measure
is defined as $Dp\equiv d^{3}{\bf p}/\left[(2\pi \hbar)^{3}2E_{{\bf p}}\right]$.
The MVSD \textit{in momentum space} is given as
$\tilde{f}_{rs}(\mathbf{p}_{1},\mathbf{p}_{2}) = \left\langle
a_{\mathbf{p}_{2},s}^{\dagger}a_{\mathbf{p}_{1},r} \right\rangle$,
where $\langle \mathcal{O} \rangle \equiv \mathrm{Tr} (\rho \,\mathcal{O})$
denotes the expectation value of the operator $\mathcal{O}$ in the ensemble
characterized by the density matrix (\ref{eq:density_matrix}).
We define the MVSD \textit{in phase space} by taking the Fourier transform with respect to
the relative momentum $q^{\mu}\equiv p_{1}^{\mu}-p_{2}^{\mu}$,
\begin{equation}
f_{rs}(x,p)\equiv\int\frac{d^{4}q}{2(2\pi \hbar)^{3}}\exp\left(-\frac{i}{\hbar}q\cdot x\right)
\delta(p\cdot q)\tilde{f}_{rs}(\mathbf{p}_{1},\mathbf{p}_{2})\;,
\label{eq:distribution_phase_space}
\end{equation}
where $p^{\mu}\equiv(p_{1}^{\mu}+p_{2}^{\mu})/2$ is the average momentum.
The MVSD (\ref{eq:distribution_phase_space}) is a generalization of the classical
distribution function to the case of quantum particles with spin 1/2, which
satisfies $f_{rs}^{\ast}=f_{sr}$, indicating that $f_{rs}(x,p)$ is a Hermitian matrix.
We note that $p_{1}^{\mu}$ and $p_{2}^{\mu}$
are restricted to the mass-shell $p_{1}^{2}=p_{2}^{2}=m^{2}$, leading to the constraint
$p\cdot q=0$.
The MVSD has previously been used to derive the equilibrium form of the polarization
\citep{Becattini:2009wh,Becattini:2013fla}. In this work, we will derive the Boltzmann equation for the MVSD
that describes its dynamical evolution in a nonequilibrium system.

For a system of weakly interacting particles, one expects that
$\tilde{f}_{rs}(\mathbf{p}_{1},\mathbf{p}_{2})$ has nonvanishing
values only when $|{\bf p}_{1}-{\bf p}_{2}| = |\mathbf{q}| \ll |\mathbf{p}|
= |{\bf p}_{1}+{\bf p}_{2}|$. As a consequence,
the gradient of the MVSD (\ref{eq:distribution_phase_space})
satisfies $\hbar\left|\boldsymbol{\nabla}_{{\bf x}}f_{rs}(x,p)\right|\ll
\left|{\bf p}f_{rs}(x,p)\right|$,
ensuring the validity of the $\hbar$ expansion, which we will employ in the following.

\paragraph*{Wigner function. --- }

We now relate the MVSD (\ref{eq:distribution_phase_space}) to the Wigner function
\cite{Heinz:1983nx,Vasak:1987um}
\begin{equation}
W(x,p)=\int\frac{d^{4}y}{(2\pi \hbar)^{4}}e^{-ip\cdot y/\hbar}
\left\langle \bar{\psi}\left(x+\frac{y}{2}\right)\otimes\psi\left(x-\frac{y}{2}\right)\right\rangle
\label{eq:definition_Wigner},
\end{equation}
where $\otimes$ denotes the Kronecker product.
Let us consider the collisionless case, i.e., we assume that the Dirac-field operators
in Eq.\ (\ref{eq:definition_Wigner}) fulfill the non-interacting Dirac equation
\cite{Itzykson:1980rh}. Inserting these operators into the definition
(\ref{eq:definition_Wigner}), performing a gradient expansion, and keeping terms of
first order in $\hbar$ we arrive at
\begin{align}
W(x,p) & \equiv\frac{1}{(2\pi)^{3}}\theta(p^{0})\delta(p^{2}-m^{2})\nonumber \\
 & \!\!\!\!\!\!\!\!\!\!\!\!\!\!\!\!\!\!\times\sum_{rs}
 \left[\bar{u}_{s}(\mathbf{p})\otimes u_{r}(\mathbf{p})
 +i\hbar\, \boldsymbol{\mathcal{U}}_{sr}({\bf p})\cdot
 \boldsymbol{\nabla}_{\mathbf{x}}\right]f_{rs}(x,p)\;.\label{eq:Wigner_first_order}
\end{align}
The momentum in Eq.\ (\ref{eq:Wigner_first_order})
is restricted to the mass-shell $p^{2}=m^{2}$, while off-shell
corrections arise at second order in $\hbar$. The matrix-valued Berry
connection for Dirac fermions is defined as \cite{Chen:2013iga,Sheng:2020oqs}
\begin{equation}
\boldsymbol{\mathcal{U}}_{sr}({\bf p})\equiv
\frac{1}{2}\! \left\{ \vphantom{\frac{}{}} \!
\left[\boldsymbol{\nabla}_{\mathbf{p}}\bar{u}_{s}(\mathbf{p})\right]
\otimes u_{r}(\mathbf{p})-\bar{u}_{s}(\mathbf{p})\otimes
\left[\boldsymbol{\nabla}_{\mathbf{p}}u_{r}(\mathbf{p})\right]\! \right\} ,
\label{eq:Berry_connection}
\end{equation}
which is a $2\times2$ matrix in spin space and a $4\times4$ matrix
in Dirac space.

We further decompose the Wigner function in terms of the generators
of the Clifford algebra,
\begin{equation}
W=\frac{1}{4}\left(\mathcal{F}+i\gamma^{5}\mathcal{P}+\gamma^{\mu}\mathcal{V}_{\mu}
+\gamma^{5}\gamma^{\mu}\mathcal{A}_{\mu}
+\frac{1}{2}\sigma^{\mu\nu}\mathcal{S}_{\mu\nu}\right)\;,
\end{equation}
where $\sigma^{\mu \nu} \equiv \frac{i}{2} [\gamma^\mu, \gamma^\nu]$.
The vector component $\mathcal{V}^{\mu}$ has a clear physical meaning:
it is the current density in phase space \cite{Vasak:1987um}. Its zeroth
component, $\mathcal{V}^{0} \equiv \text{Tr}(\gamma^{0}W)$, where ``Tr''
denotes the trace in Dirac space, is given by
\begin{equation} \label{eq:vector_0}
\mathcal{V}^{0}=\frac{2}{(2\pi \hbar)^{3}}\theta(p^{0})\delta(p^{2}-m^{2})
E_{{\bf p}}\, \text{tr}\left[F(x,p)\right]\;,
\end{equation}
where ``tr'' denotes the trace in spin space and
\begin{align}
F(x,p) & \equiv f(x,p)\nonumber \\
 & \!\!\!\!\!\!\!\!\!\!\!\!\!\!\!\!\!\!\!\!\!+\frac{\hbar}{4(u_{0}\cdot p)(u_{0}\cdot p+m)}
 \epsilon^{\mu\nu\alpha\beta}u_{0,\mu}p_{\nu}\left\{ n_{\beta}({\bf p}),\,
\partial_{\alpha} f(x,p)\right\}\;. \label{eq:frame dependent MVDF}
\end{align}
Here, $\{ A, B \} \equiv AB + BA$ for two arbitrary $2 \times 2$ matrices $A,B$.
The vector $u_{0}^{\mu}\equiv(1,0,0,0)$ defines the rest frame
of a specific system, called ``laboratory system'' in the following.
The quantity $\text{tr}\left[F(x,p)\right]$ in Eq.\ (\ref{eq:vector_0}) is
therefore the particle number density observed in the laboratory frame.
The spin-polarization vector $n^{\mu}({\bf p})$ is
a $2 \times 2$ matrix and is defined as
\begin{align}
n_{sr}^{\mu}({\bf p}) & \equiv \frac{1}{2m} \bar{u}_s(\mathbf{p})
\gamma^\mu \gamma_5 u_r(\mathbf{p}) \nonumber \\
& = \left(\frac{\mathbf{p}\cdot\boldsymbol{\tau}_{sr}}{m},\,
\boldsymbol{\tau}_{sr}+\frac{\mathbf{p}\cdot\boldsymbol{\tau}_{sr}}{m(E_{\mathbf{p}}+m)}
\mathbf{p}\right)\;, \label{eq:spin-polarization_vector}
\end{align}
where $\boldsymbol{\tau}_{sr} = \chi_s^\dagger \boldsymbol{\sigma} \chi_r$,
with $\boldsymbol{\sigma}$ being the vector of Pauli matrices and
$\chi_r^\dagger, \chi_s$ the Pauli spinors.

The gradient term in Eq.\ (\ref{eq:frame dependent MVDF}) arises from
the Berry connection (\ref{eq:Berry_connection}). It can be
absorbed into $f(x,p)$ by introducing a matrix-valued shift
$\delta x$ in space-time and defining the Taylor expansion up to first order in $\delta x$ as
$f(x+\delta x)= f(x)+\left\{ \delta x^{\mu},\partial_{\mu}f(x)\right\} /2 + \mathcal{O}(\delta x^2)$.
The new position $x+\delta x$ agrees with the canonical position
operator proposed in Refs.\ \cite{Pryce:1948pf,Bacry:1988jn,Rivas:2002bk,Lorce:2021gxs},
which is interpreted as the energy center for a particle with spin.
Then, we identify $\text{tr}\left[F(x,p)\right]$ as the particle number density for particles
with momentum $\mathbf{p}$ and energy center at $x+\delta x$ in the laboratory frame.

The vector and axial-vector components of the Wigner function
can be expressed in terms of $F(x,p)$ as
\begin{eqnarray}
\mathcal{V}^{\mu} & = & C\,\text{tr}\left[ \left(p^{\mu}+
\hbar S_{u_0}^{\mu \nu}\partial_{\nu}\right)F\right] \;,\label{eq:current} \\
\mathcal{A}^{\mu} & = & C\,\text{tr}\left\{ \left[mn^{\mu}({\bf p})
+\hbar L_{u_0}^{\mu \nu} \partial_{\nu}\right]F\right\} \;, \label{eq:spin}
\end{eqnarray}
where the prefactor $C=\left[2/(2\pi \hbar)^{3}\right]\theta(p^{0})\delta(p^{2}-m^{2})$
and
\begin{align}
&S_{u_0}^{\mu\nu} \equiv \frac{m}{2(u_0\cdot p)}\epsilon^{\mu\nu\alpha\beta}
n_{\alpha}({\bf p})u_{0,\beta}\;, \label{eq:spin_tensor} \\
&L_{u_0}^{\mu \nu} \equiv \frac{1}{2(u_0\cdot p)}\epsilon^{\mu\nu\alpha\beta}
p_{\alpha}u_{0,\beta}\;. \label{eq:L}
\end{align}
Equations (\ref{eq:current}) and (\ref{eq:spin}) agree with the results derived in Appendix C of
Ref.~\cite{Hattori:2019ahi}.
In the massless case, the spin-polarization vector aligns (anti-aligns) with the momentum
for positive (negative) helicity, i.e., one has to replace
$ m n^\mu({\bf p})/2 \rightarrow \lambda p^\mu$, where
$\lambda = \pm 1/2$, and the expression (\ref{eq:spin_tensor}) agrees with the
spin tensor introduced in Eq.\ (3) of Ref.\ \cite{Chen:2015gta}. The
vector current (\ref{eq:current}) has a rather similar form
as the result in the massless case, cf.\ Eq.\ (7) of Ref.\ \cite{Chen:2015gta}.

In order to clarify the physical meaning of the terms in Eqs.\ (\ref{eq:current}) and (\ref{eq:spin}),
we first consider the canonical angular-momentum tensor
\begin{equation}
J^{\lambda \mu\nu}=x^\mu T^{\lambda\nu}-x^\nu T^{\lambda\mu}+\hbar S^{\lambda\mu\nu}\;,
\end{equation}
where the canonical energy-momentum tensor and the canonical spin angular-momentum tensor are defined
as $T^{\mu\nu}\equiv\int d^{4}p\,p^{\mu}\mathcal{V}^{\nu}$ and
$S^{\lambda\mu\nu}\equiv-\int d^{4}p\,\epsilon^{\lambda\mu\nu\alpha}
\mathcal{A}_{\alpha}/2$, respectively. The corresponding conserved charge $J^{0 \mu\nu}$ can be calculated by
substituting Eqs.\ (\ref{eq:current}) and (\ref{eq:spin}) into $J^{\lambda\mu\nu}$ and then take the $\lambda=0$
component. At leading order in $\hbar$, the result reads
\begin{equation}
J^{0 \mu\nu}= \int \frac{d^{3}\bf{p}}{(2\pi \hbar)^3}\,\text{tr}[(x^\mu p^\nu-x^\nu p^\mu +\hbar S_{u_0}^{\mu\nu})F]\;,
\label{eq:Conserved_AM}
\end{equation}
We therefore identify $S_{u_0}^{\mu\nu}$ defined in Eq.\ (\ref{eq:spin_tensor}) as the rank-2 spin tensor in the
laboratory frame and the term $\sim \text{tr}(S_{u_0}^{\mu\nu} \partial_\nu F)$ in Eq.\ (\ref{eq:current}) as the
magnetization current induced by the inhomogeneity of the distribution $F$.

\begin{figure}
\includegraphics[width=6cm]{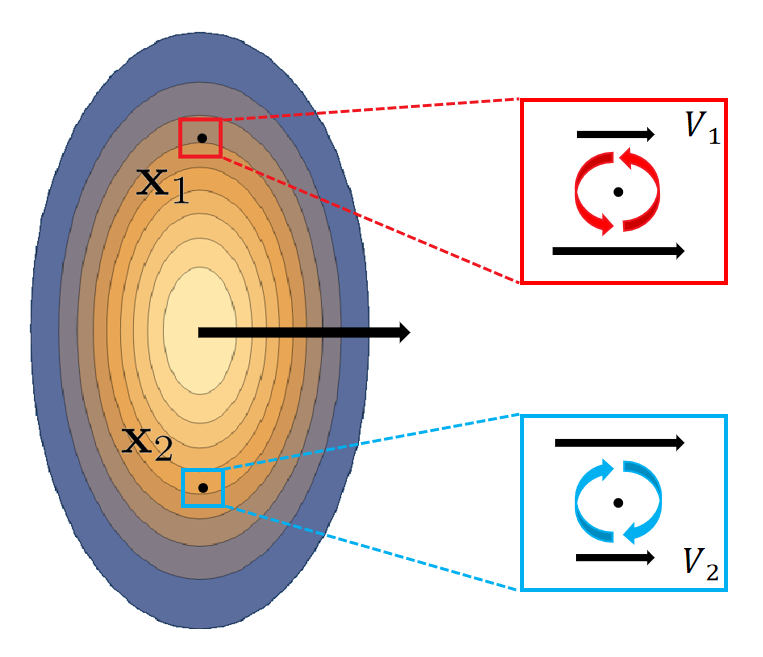}

\caption{Local angular momentum density generated by an inhomogeneous current density. \label{fig:OAM_momentum}}

\end{figure}

On the other hand, $\mathcal{A}_\mu$ in Eq.\ (\ref{eq:spin}) is
interpreted as the spin angular-momentum density, which consists of two parts. The first part is an intrinsic
spin density $\sim \text{tr}\left[n^{\mu}({\bf p})F\right]$, which is proportional to the polarization vector multiplied with the
distribution function. The second part is a motion-induced part $\sim \text{tr}(L_{u_0}^{\mu\nu}\partial_\nu F)$, which is
generated through the spin-orbit coupling. Considering a Gaussian-type particle number-density distribution moving to the
right with momentum $\mathbf{p}$ at time $t=0$, cf.\ Fig.~\ref{fig:OAM_momentum}, the local current density is then
given by ${\bf p}\, \text{tr}\left[F(\mathbf{x},p)\right]$. For the region $V_1$ in the vicinity of the point ${\bf x}_{1}$,
the current density nearer to the center of the distribution is larger than that further away
from the center, leading to a nonvanishing OAM
\begin{align}
{\bf L}_{V_1}&=\int_{V_1} d^{3}{\bf x}^{\prime}
\left({\bf x}^{\prime}-{\bf x}_{1}\right)\times{\bf p}\,
\text{tr}\left[F({\bf x}^{\prime},p)\right] \nonumber\\
&\simeq -\frac{1}{3} {\bf p}\times\boldsymbol{\nabla}_{{\bf x}_{1}}\text{tr}[F({\bf x}_{1},p)]
\int_{V_1} d^{3}{\bf x}^{\prime} \left({\bf x}^{\prime}-{\bf x}_{1}\right)^2\;, \label{eq:L_V1}
\end{align}
where we have made a gradient expansion for $F({\bf x}^\prime,p)$  near ${\bf x}_1$. By comparing $L_{V_1}$ with
the OAM of a local vortex with kinetic vorticity $\boldsymbol{\omega}$,
\begin{equation}
{\bf L}_{\boldsymbol{\omega}}= \frac{1}{3}{\boldsymbol{\omega}}E_{\bf p} \text{tr}[F({\bf x}_{1},p)]
\int_{V_1} d^{3}{\bf x}^{\prime} \left({\bf x}^{\prime}-{\bf x}_{1}\right)^2\;,
\end{equation}
one can find that the OAM generated by the inhomogeneous current density near point ${\bf x}_1$ corresponds to that
of an anti-clockwise rotating vortex with
\begin{equation}
{\boldsymbol \omega} = -\frac{{\bf p}\times\boldsymbol{\nabla}_{{\bf x}_{1}}\text{tr}[F({\bf x}_{1},p)]}{E_{\bf p}
\text{tr}[F({\bf x}_{1},p)]}\;.
\end{equation}
Similarly, the OAM near the point ${\bf x}_{2}$ is equivalent
to the contribution of a clockwise rotating vortex, because the
density gradient at $\mathbf{x}_2$
points in the opposite direction than that at point ${\bf x}_{1}$. Through the spin-orbit coupling, the OAM results in a
nonvanishing spin density, i.e., the term $\sim \text{tr}(L_{u_0}^{\mu\nu}\partial_\nu F)$ in Eq.\ (\ref{eq:spin}).

\paragraph*{Lorentz transformation. ---}

We now study how the MVSD transforms under a Lorentz boost from the laboratory frame,
characterized by the frame vector $u_0^\mu=(1,0,0,0)$,
to a new reference frame called ``$u$-frame'' in the following, which moves with velocity
$u^{\mu}=(\gamma,\gamma{\bf v})$ with respect to the laboratory frame.
This Lorentz boost is denoted as $\Lambda_{uu_{0}}$ (without specifying the
particular representation of the Lorentz group that this boost acts on).

The plane-wave state introduced in Eq.\ (\ref{eq:plane_wave_state}) transforms as
\begin{equation}
U(\Lambda_{uu_{0}})\left|{\bf p},s\right\rangle =\sum_{r}\left|{\bf p}^{\prime},r\right\rangle
D_{rs}(R_{u,{\bf p}})\;, \label{eq:Wigner_rotation}
\end{equation}
where $U(\Lambda_{uu_{0}})$ is a unitary representation for $\Lambda_{uu_{0}}$
(appropriate for acting on the Fock-space state $\left|{\bf p},s\right\rangle$)
and ${\bf p}^{\prime}$ is the spatial component of the momentum in
the $u$-frame, satisfing
$p^{\prime\mu}=\left(\Lambda_{uu_{0}}\right)_{\ \nu}^{\mu}p^{\nu}$.
Note that in general a Lorentz boost changes the spin of a particle, which
is also known as Wigner rotation $R_{u,{\bf p}}$. The latter is defined as the product
of three Lorentz boosts, $R_{u,{\bf p}}\equiv\Lambda_{\text{rest},u}\Lambda_{uu_{0}}
\Lambda_{u_{0},\text{rest}}$, where ``rest'' denotes the rest frame with
$p_{\text{rest}}^{\mu}=\left(\Lambda_{\text{rest},u_{0}}\right)_{\ \nu}^{\mu}p^{\nu} =(m,{\bf 0})$.
In Eq.\ (\ref{eq:Wigner_rotation}), the Wigner rotation
is encoded in the unitary $2 \times 2$ matrix $D_{rs}(R_{u,{\bf p}})$ in spin space.
For massless particles, this matrix is diagonal and reduces to a mere phase
factor.

The transformation behavior of the Dirac spinors follows from Eq.\ (\ref{eq:Wigner_rotation}),
\begin{equation}
\left(\Lambda_{uu_{0}}\right)_{\frac{1}{2}}u_{s}({\bf p})=\sum_r u_r({\bf p}^{\prime})
D_{rs}(R_{u,{\bf p}})\;,\label{eq:transformation_spinors}
\end{equation}
which gives
\begin{equation}
D_{rs}(R_{u,{\bf p}})=\frac{1}{2m}\bar{u}_{r}({\bf p}^{\prime})
\left(\Lambda_{uu_{0}}\right)_{\frac{1}{2}}u_{s}({\bf p})\;,\label{eq:transform_matrix}
\end{equation}
where $\left(\Lambda_{uu_{0}}\right)_{\frac{1}{2}}$
is the spinor
representation of the Lorentz boost $\Lambda_{uu_{0}}$.

Demanding that the density matrix in Eq.\ (\ref{eq:density_matrix})
transforms as $\rho^{\prime}=U(\Lambda_{uu_{0}})\rho\, U^{\dagger}(\Lambda_{uu_{0}})$
and then using Eq.\ (\ref{eq:Wigner_rotation}), we obtain the transformation
property of the MVSD in momentum space,
\begin{equation}
\tilde{f}^{\prime}(\mathbf{p}_{1}^{\prime},\mathbf{p}_{2}^{\prime})
=D(R_{u,{\bf p}_{1}})\tilde{f}(\mathbf{p}_{1},\mathbf{p}_{2})D^{\dagger}(R_{u,{\bf p}_{2}})\;,
\end{equation}
where $\tilde{f}^{\prime}$, $\mathbf{p}_{1}^{\prime}$, and $\mathbf{p}_{2}^{\prime}$
are quantities in the $u$-frame, with
$p_{i}^{\prime\mu}=\left(\Lambda_{uu_{0}}\right)_{\ \nu}^{\mu}p_{i}^{\nu}$
for $i=1,2$. Here, both $D$ and $\tilde{f}$ are $2\times2$ matrices,
whose indices are omitted for the sake of simplicity. Substituting the above
relation into Eq.\ (\ref{eq:distribution_phase_space}), we derive
\begin{align}
f^{\prime}(x^{\prime},p^{\prime}) & =D(R_{u,{\bf p}})f(x,p)D^{\dagger}(R_{u,{\bf p}})
\nonumber \\
 & \!\!\!\!\!\!\!\!\!\!\!\!-\frac{i\hbar}{2}\left[\boldsymbol{\nabla}_{{\bf p}}D(R_{u,{\bf p}})\right]
 \cdot\left[\boldsymbol{\nabla}_{{\bf x}}f(x,p)\right]D^{\dagger}(R_{u,{\bf p}})\nonumber \\
 & \!\!\!\!\!\!\!\!\!\!\!\!+\frac{i\hbar}{2}D(R_{u,{\bf p}})\left[f(x,p)
 \overleftarrow{\boldsymbol{\nabla}}_{{\bf x}}\right]\cdot\left[\boldsymbol{\nabla}_{{\bf p}}
 D^{\dagger}(R_{u,{\bf p}})\right]\;.\label{eq:transform_distribution_phase_space}
\end{align}
Given explicit expressions for the Dirac spinors, we can derive
the matrix $D$ in Eq.\ (\ref{eq:transform_matrix}).
Then, we substitute $D$ into Eq.\ (\ref{eq:transform_distribution_phase_space}).
The Lorentz transform of $F$, defined in Eq.\ (\ref{eq:frame dependent MVDF}),
can be expressed with the help of Eq.\ (\ref{eq:transform_distribution_phase_space}) as
\begin{equation}
F^{\prime}=D(R_{u,{\bf p}})\left[F+\frac{\hbar}{2}\left\{ \Delta_{u_0u}^{\mu},\partial_{\mu}F\right\} \right]D^{\dagger}(R_{u,{\bf p}})\;,\label{eq:transform_distribution}
\end{equation}
where
\begin{equation} \label{eq:Delta_uu0}
\Delta_{u_0u}^{\mu}\equiv\Delta_{u_{0}}^{\mu}-\Delta_{u}^{\mu}\;.
\end{equation}
with the frame-dependent shift term being
\begin{equation}
\Delta_{u}^{\mu}\equiv-\frac{1}{m^2} S_u^{\mu \nu}p_\nu
\equiv \frac{1}{m} L_u^{\mu \nu}n_\nu(\mathbf{p})\;. \label{eq:shift_term}
\end{equation}
Note that, if the $u$-frame is the rest frame of the particle,
$u_{\mathrm{rest}}^\mu = p^\mu/m$, we have
$\Delta_{u_{\mathrm{rest}}}^\mu =0$ and $\Delta_{u_0u_{\mathrm{rest}}}^{\mu}
= \Delta_{u_0}^\mu$.
Equation (\ref{eq:transform_distribution}) states that, under a Lorentz boost,
the transformed $F$
contains a rotation in spin space and an additional contribution from
the gradient of $F$. The latter part arises because the magnetization
current induced by an inhomogeneous spin angular momentum density
and the OAM induced by an inhomogeneous current density
are frame-dependent quantities. The shift term (\ref{eq:shift_term}) has been studied
for non-relativistic electron scattering off a central potential in
ferromagnets in Ref.\ \cite{Berger:1970}, and in this work we generalize
it to the relativistic case.

We now discuss the behavior of $\mathcal{V}^{\mu}$
and $\mathcal{A}^{\mu}$ in Eqs.\ (\ref{eq:current}) and (\ref{eq:spin}) under
Lorentz transformations. In the $u$-frame, the vector current reads
\begin{equation} \label{eq:V_LT}
\mathcal{V}^{\prime \mu} = C\, \text{tr} \left[ \left( p^{\prime \mu}
+ \hbar S_{u_0}^{\prime \mu \nu}
\partial^\prime_\nu \right) F^\prime (x^\prime,p^\prime) \right]\;,
\end{equation}
where $S_{u_0}^{\prime \mu \nu}$ is the spin tensor (\ref{eq:spin_tensor}) with
$n^\mu(\mathbf{p})$ and $p^\mu$ replaced by
$n^{\prime \mu}(\mathbf{p}^\prime)$ and $p^{\prime \mu}$,
respectively, while the frame vector $u_0^\mu=(1,0,0,0)$ is not transformed
due to the frame dependence of $S_{u_0}^{\prime \mu \nu}$.
We now replace all $u$-frame Lorentz tensors in Eq.\ (\ref{eq:V_LT})
by laboratory-frame tensors, using
$a^{\prime \mu} = (\Lambda_{uu_0})^\mu_{\ \nu} a^\nu$
and $u_0^\mu=(\Lambda_{uu_0})^\mu_{\ \nu} u^\nu$.
The Lorentz transformation of the spin-polarization vector can be deduced
from its definition (\ref{eq:spin-polarization_vector})
and the transformation behavior
of the Dirac spinors (\ref{eq:transformation_spinors}),
\begin{equation} \label{eq:spin-polarization_LT}
n^{\prime \mu}({\bf p}^\prime) = D(R_{u,\mathbf{p}}) (\Lambda_{uu_0})^\mu_{\ \nu}
n^\nu(\mathbf{p}) D^\dagger(R_{u,\mathbf{p}}) \;.
\end{equation}
Then, using Eq.\ (\ref{eq:transform_distribution}), the Schouten identity
for $p^\lambda \epsilon^{\mu \nu \alpha \beta}$ \cite{Sheng:2020oqs},
as well as $p \cdot n(\mathbf{p}) = 0$,
up to first order in $\hbar$ Eq.\ (\ref{eq:V_LT}) reads
\begin{equation}
\mathcal{V}^{\prime \mu}  = (\Lambda_{uu_0})^\mu_{\ \nu}
\left(\mathcal{V}^\nu +\hbar\,C \,
\text{tr} \left\{ \Delta_{u_0u}^{\nu}  p \cdot \partial F \right\}\right)\;. \label{eq:V_LT_2}
\end{equation}
To lowest order in $\hbar$, the distribution function $F$ fulfills a Boltzmann equation of the form
$p \cdot \partial F = \mathcal{C}[F]$, cf.\ Eq.\ (\ref{eq:Boltzmann_transform}) below. Then, in the
absence of collisions, $\mathcal{C}[F]=0$, the second term in Eq.\ (\ref{eq:V_LT_2}) vanishes, so the vector current transforms in a Lorentz-covariant manner
\cite{Chen:2015gta,Hattori:2019ahi}.

In order to ensure the covariance of $\mathcal{V}^\mu$ in the presence of collisions,
an additional term
\begin{equation}
\delta\mathcal{V}^{\mu}
=\hbar\, C\,\text{tr}\left(\Delta_{u_{0}\bar{u}}^{\mu} p \cdot \partial F\right)
\end{equation}
needs to be added to $\mathcal{V}^{\mu}$ in Eq.\ (\ref{eq:current}), where
$\bar{u}^\mu$ is an arbitrary frame vector and $\Delta^\mu_{\bar{u}}$,
cf.\ Eq.\ (\ref{eq:shift_term}), transforms as a Lorentz vector. Then, defining
$\hat{\mathcal{V}}^{\mu}\equiv \mathcal{V}^{\mu} + \delta\mathcal{V}^{\mu}$
one can show that $\hat{\mathcal{V}}^{\prime \mu} = (\Lambda_{uu_0})^\mu_{\ \nu}
\hat{\mathcal{V}}^{\nu}$ up to order $\mathcal{O}(\hbar)$, i.e.,
$\hat{\mathcal{V}}^{\mu}$ transforms as a Lorentz vector.

To prove the Lorentz covariance of the axial-vector current, an analogous calculation
as in the vector case yields~\cite{Hattori:2019ahi}
\begin{equation} \label{eq:AA}
\mathcal{A}^{\prime \mu}  = (\Lambda_{uu_0})^\mu_{\ \nu} \,
\mathcal{A}^\nu\;,
\end{equation}
using Eqs.\ (\ref{eq:shift_term}), (\ref{eq:spin-polarization_LT}),
as well as the relation
\begin{equation} \label{eq:nn}
\left\{ n^\mu (\mathbf{p}), n^\nu(\mathbf{p}) \right\}
= - 2 \left( g^{\mu \nu} - \frac{p^\mu p^\nu}{m^2} \right)\;.
\end{equation}
Equation (\ref{eq:AA}) shows that the axial-vector current always transforms in a Lorentz-covariant manner,
even in the case with collisions.

\paragraph{Collision term. --- }

Now we consider the case of binary elastic collisions.
In the classical case, all incoming or outgoing particles have well-defined positions
and the scattering process happens at one space-time point $x$.
However, for quantum particles the position has a finite uncertainty and therefore particles can interact with each other
over a finite distance. Such a nonlocal collision involves a finite OAM, which can be converted into the particle's spin,
or vice versa.

In the laboratory frame, the conserved angular-momentum tensor relative to a specific point $x_0^\mu$ is given by
$J^{0\mu\nu}$,
\begin{align}
J^{0 \mu\nu}=&\int\frac{d^3{\bf p}}{(2\pi\hbar)^3}\text{tr}\left\{\left[(x^{\mu}-x_0^{\mu}) p^{\nu}\right.\right. \nonumber \\
&\left.\left.-(x^{\nu}-x_0^{\nu}) p^{\mu}+\hbar S_{u_0}^{\mu\nu}\right]F(x ,p)\right\}\;.
\end{align}
One can clearly identify the last term as the contribution from spin angular momentum and the remaining terms as the
OAM part. However, note that the following identity holds up to order $\hbar$,
\begin{equation}
\Delta^\mu_{u_0u}p^\nu-\Delta^\nu_{u_0u}p^\mu+S^{\mu\nu}_{u_0}-S^{\mu\nu}_u=0\;,
\end{equation}
which can be proved by employing Eqs.\ (\ref{eq:Delta_uu0}), (\ref{eq:shift_term}), and the Schouten identity.
Here $S_{u}^{\mu\nu}$ is the spin tensor defined in Eq.\ (\ref{eq:spin_tensor}) with
$u_0^\mu$ replaced by $u^\mu$. Then it is possible to express $J^{0\mu\nu}$ in another form with the help of the
frame vector $u^\mu$,
\begin{align}\label{eq:conserved_AM}
J^{0 \mu\nu}=&\int\frac{d^3{\bf p}}{(2\pi\hbar)^3}\text{tr}\left\{\left[(x^{\mu}-x_0^{\mu}
-\hbar\Delta^\mu_{u_0u}) p^{\nu}\right.\right. \nonumber \\
&\left.\left.-(x^{\nu}-x_0^{\nu}-\hbar\Delta^\nu_{u_0u}) p^{\mu}+\hbar S_{u}^{\mu\nu}\right]F(x ,p)\right\}\;.
\end{align}
Using the Lorentz transform (\ref{eq:transform_distribution}) of the MVSD, we can prove that the
last term in Eq.\ (\ref{eq:conserved_AM}) is related to the spin angular momentum in the reference frame moving with
velocity $u^\mu=(\gamma, \gamma{\bf v})$ relative to the laboratory frame,
\begin{equation}
\text{tr}[\hbar S^{\mu\nu}_{u}F(x,p)]
=(\Lambda_{u_0u})^\mu_{\ \alpha}(\Lambda_{u_0u})^\nu_{\ \beta}\text{tr}[\hbar
S^{\prime\alpha\beta}_{u_0}F^\prime(x^\prime,p^\prime)]\;,
\end{equation}
where we have dropped terms of order $\mathcal{O}(\hbar^2)$. We then conclude that the conserved
angular momentum $J^{0\mu\nu}$ is independent of the reference frame, while the decomposition of
$J^{0\mu\nu}$ into spin and OAM depends on the choice of $u^\mu$.

Assuming that, at leading order in $\hbar$, the MVSD satisfies a Boltzmann equation of the same form
as in the classical case, $p\cdot\partial F(x,p)=\mathcal{C}[F]+\mathcal{O}(\hbar)$, one immediately obtains the
requirement for angular-momentum conservation during collisions,
\begin{align}
\int& Dp\,\text{tr}\left\{\left[(x^{\mu}-x_0^{\mu}-\hbar\Delta^\mu_{u_0u}) p^{\nu}\right.\right. \nonumber \\
&\left.\left.-(x^{\nu}-x_0^{\nu}-\hbar\Delta^\nu_{u_0u}) p^{\mu}+\hbar S_{u}^{\mu\nu}\right]\mathcal{C}[F]\right\}=0\;.
\end{align}
In the massless case, there exists a special frame, the so-called ``no-jump'' frame
\cite{Chen:2015gta}, where the incoming particles collide at the same position $x$.
In our case of massive particles, the analogue is a frame,
characterized by a frame vector $\bar{u}^\mu$,
where the spin angular momentum and thus, by conservation of total angular momentum, also
the OAM are separately conserved, i.e.,
\begin{equation}\label{constraint_ubar}
\hbar\int Dp\,\text{tr}\left(S_{\bar{u}}^{\mu\nu}\mathcal{C}[F]\right) =0\;.
\end{equation}
In this frame, the Boltzmann equation is assumed to take the same
form as in the classical case,
\begin{equation}
\bar{p}\cdot \bar{\partial} \bar{F}(\bar{x},\bar{p}) = \bar{\mathcal{C}}[\bar{F}]\;.
\end{equation}
Then, using Eq.\ (\ref{eq:transform_distribution}) we conclude that in the laboratory frame
the Boltzmann equation reads
\begin{equation} \label{eq:Boltzmann_transform}
p\cdot\partial\left[F(x,p)+\frac{\hbar}{2}\left\{ \Delta_{u_{0}\bar{u}}^{\mu},\partial_{\mu}
F(x,p)\right\} \right]=\mathcal{C}[F]\;,
\end{equation}
where the collision term $\mathcal{C}[F]$ is related to $\bar{\mathcal{C}}[\bar{F}]$ as
\begin{equation}
\bar{\mathcal{C}}[\bar{F}]=D(R_{\bar{u},{\bf p}})\mathcal{C}[F]
D^{\dagger}(R_{\bar{u},{\bf p}}) + \mathcal{O}(\hbar)\;.
\label{eq:C_transform}
\end{equation}
Since the left-hand side of Eq.\ (\ref{eq:Boltzmann_transform}) is Hermitian,
the collision term must also be Hermitian, $\mathcal{C}^{\dagger}[F]=\mathcal{C}[F]$.
In the $\bar{u}$-frame, collisions are local, and the collision term
$\bar{\mathcal{C}}[\bar{F}]$ has the same form as
in the classical case. One can then show that, in the laboratory frame,
\begin{eqnarray}
\mathcal{C}_{rs}[F] & = & \frac{1}{4}\sum_{r_{i},s_{i}}\int DP
\mathcal{M}\left(p_{1},p_{2};s_{1},s_{2}\rightarrow p,p_{3};r_{0},r_{3}\right)
\nonumber \\
 &  & \times\mathcal{M}^{\ast}\left(p_{1},p_{2};r_{1},r_{2}\rightarrow
 p,p_{3};r,s_{3}\right)\nonumber \\
 &  & \times\Biggl\{\tilde{F}_{r_{1}s_{1}}(p_{1})\tilde{F}_{r_{2}s_{2}}(p_{2})
 \left[\delta_{r_{0}s}-\tilde{F}_{r_{0}s}(p)\right]\nonumber \\
 &  & \times\left[\delta_{r_{3}s_{3}}-\tilde{F}_{r_{3}s_{3}}(p_{3})\right]
 -\tilde{F}_{r_{0}s}(p)\tilde{F}_{r_{3}s_{3}}(p_{3})\nonumber \\
 &  & \times\left[\delta_{r_{1}s_{1}}-\tilde{F}_{r_{1}s_{1}}(p_{1})\right]
 \left[\delta_{r_{2}s_{2}}-\tilde{F}_{r_{2}s_{2}}(p_{2})\right]\Biggr\}\nonumber \\
 &  & +\text{h.c.}\;, \label{eq:collisionterm_nojump}
\end{eqnarray}
where the invariant integration measure $DP\equiv Dp_{1}Dp_{2}Dp_{3}\,
(2\pi\hbar)^{4}\delta^{(4)}(p_{1}+p_{2}-p-p_{3})$ and ``h.c'' stands for the Hermitian conjugate (complex conjugate and interchanging $r$ and $s$) of the first term.
In Eq.\ (\ref{eq:collisionterm_nojump}), the distribution function $\tilde{F}$ is defined as
\begin{equation} \label{eq:tildeF}
\tilde{F}(x,p)\equiv F(x,p)+\frac{\hbar}{2}
\left\{ \Delta_{u_{0}\bar{u}}^{\mu},\partial_{\mu}F(x,p)\right\}\;,
\end{equation}
and we suppressed the $x$-dependence of $\tilde{F}$ for the sake of simplicity.
Under a Lorentz boost the transition amplitude transforms as
\begin{align}
\bar{\mathcal{M}} & =\sum_{s_{0}s_{1}s_{2}s_{3}}D_{s_{0}\bar{s}_{0}}
D_{s_{3}\bar{s}_{3}}D_{\bar{s}_{1}s_{1}}^{\dagger}D_{\bar{s}_{2}s_{2}}^{\dagger}
\nonumber \\
 & \times\mathcal{M}\left(p_{1},p_{2};s_{1},s_{2}\rightarrow p,p_{3};s_{0},s_{3}\right)\;.
\label{eq:M_prime}
\end{align}
The Wigner rotation matrices in this equation partially cancel with those for the
MVSDs in Eq.\ (\ref{eq:transform_distribution}), which ensures that $\mathcal{C}[F]$
transforms as in Eq.\ (\ref{eq:C_transform}).
Note that the spin-orbit coupling enters the collision term through
the presence of the shift term $\Delta_{u_{0}\bar{u}}^{\mu}$ in the
definition of $\tilde{F}$, making
the collision term nonlocal at first order in $\hbar$, cf.\ Refs.\ \cite{Weickgenannt:2020aaf,Weickgenannt:2021cuo,Sheng:2021kfc}.

One can further check that the Boltzmann equation (\ref{eq:Boltzmann_transform}) fulfills the local conservation law for
total angular momentum,
\begin{align}
&\hbar\partial_{\lambda}S^{\lambda\mu\nu}+T^{\mu\nu}-T^{\nu\mu} \nonumber \\
&=-2\hbar \int Dp\,\text{tr}\left(S_{\bar{u}}^{\mu\nu}\mathcal{C}[F]\right)
+ \mathcal{O}(\hbar^2)=0\;,
\label{eq:local_AM_conservation}
\end{align}
where $S^{\lambda\mu\nu}$ is the canonical spin angular-momentum tensor defined above,
$T^{\mu\nu}\equiv\int d^{4}p\,p^{\mu}\hat{\mathcal{V}}^{\nu}$,
and $S_{\bar{u}}^{\mu\nu}$ is the spin tensor defined in Eq.\ (\ref{eq:spin_tensor}) with
$u_0^\mu$ replaced by $\bar{u}^\mu$.
We emphasize that in the last line in Eq.\ (\ref{eq:local_AM_conservation}) we have used Eq.\ (\ref{constraint_ubar}),
which demands that the spin is conserved in collisions in the $\bar{u}$-frame.

\paragraph{Local thermodynamical equilibrium. ---}

Usually, local thermodynamical equilibrium is defined by demanding that the collision
term vanishes. This requirement leads to the solution $\tilde{F}(x,p)=f_{FD}(x,p;\delta E)$,
with the Fermi-Dirac distribution defined as
\begin{equation}
f_{FD}(x,p;\delta E)\equiv\left\{ 1+\exp[(u\cdot p+\delta E-\mu)/T]\right\}^{-1}\;,
\end{equation}
where the energy shift $\delta E\equiv \hbar S^{\mu\nu}_{\bar{u}}\Omega_{\mu\nu}$,
with $\Omega_{\mu\nu}$ being the spin potential.
Using Eq.\ (\ref{eq:tildeF}) the MVSD in the lab frame is then
up to order $\mathcal{O}(\hbar)$ given by
\begin{equation}
F(x,p) =f_{FD}(x,p;\delta E)-\frac{\hbar}{2}\left\{ \Delta_{u_0\bar{u}}^{\mu},
\partial_{\mu}f_{FD}(x,p;\delta E)\right\} \;.
\end{equation}
With Eqs.\ (\ref{eq:spin}), (\ref{eq:L}), (\ref{eq:Delta_uu0}), and (\ref{eq:nn})
one then computes the axial-vector component of the Wigner function as
\begin{align}
\mathcal{A}^{\mu} & =C\,\text{tr}\left\{ \left[mn^{\mu}({\bf p})\, \frac{}{}
\right.\right.\nonumber \\
 & \left.\left.-\frac{\hbar}{2(\bar{u}\cdot p)}\epsilon^{\mu\nu\alpha\beta}\bar{u}_{\nu}
 p_{\alpha}\partial_{\beta}\right]f_{FD}(x,p;\delta E)\right\}\;,\label{eq:axial_current}
\end{align}
which agrees with the result that includes the thermal-shear contribution \cite{Wang:2020pej,Becattini:2021suc,Liu:2021uhn,Fu:2021pok,Becattini:2021iol,Yi:2021ryh,Buzzegoli:2021wlg,Liu:2021nyg}.
In the massless limit, Eq.\ (\ref{eq:axial_current}) smoothly reduces
to the result of CKT \cite{Chen:2015gta,Hidaka:2017auj}, which can be proved by
replacing $mn^{\mu}({\bf p})\rightarrow p^{\mu}$.

\paragraph*{Conclusions. ---}

In this work, we have derived a matrix-valued spin-dependent distribution
function (MVSD) $F(x,p)$ for quantum particles, which describes
the particle number density and intrinsic spin density
in phase space. A physical interpretation of the MVSD is
provided by expressing the vector and axial-vector
components of the Wigner function in terms of $F(x,p)$. In an inhomogeneous system, the
magnetization current and the OAM contained in an inhomogeneous momentum
distribution result in nontrivial Lorentz-transformation properties for $F(x,p)$:
in addition to the ordinary Wigner rotation, $F(x,p)$
undergoes a matrix-valued shift $\Delta_{u_{0}u}^{\mu}$ in space-time. This
term ensures that the axial-vector component, and in the collisionless case,
also the vector component of
the Wigner function transform in a Lorentz-covariant manner.
Including collisions, the vector component requires an additional contribution
to preserve Lorentz covariance.
Assuming the existence of a $\bar{u}$-frame where spin is a collisional invariant,
and using the Lorentz transformation properties
of the MVSD, we further constructed a manifestly covariant
Boltzmann equation including nonlocal terms, which give
rise to spin-orbit coupling during collisions.
In the $\bar{u}$--frame, we derive a local-equilibrium solution for
$F(x,p)$ and for the axial-vector component $\mathcal{A}^{\mu}$.
The Lorentz-covariant Boltzmann equation derived in this work provides a solid
foundation for studying spin dynamics in heavy-ion collisions and, ultimately, the
long-sought means to solve the sign problem of the longitudinal polarization.

\vspace{0.5cm}

\paragraph*{Acknowledgments: }
The authors would like to thank Yu-Chen Liu for stimulating discussions
and collaboration in the early stages of this work.
X.-L.S.\ is supported by the National Natural Science Foundation of
China (NSFC) under Grant No.\ 12047528. Q.W.\ is supported by in part by the National
Natural Science Foundation of China (NSFC) under Grants No.\ 12135011, 11890713,
12047502, and by the Strategic Priority Research Program of the Chinese Academy
of Sciences (CAS) under Grant No.\ XDB34030102.
D.H.R.\ is supported by German Research Foundation (DFG)
through the Collaborative Research Center TransRegio
CRC-TR 211 ``Strong- interaction matter under extreme conditions'' -- project number
315477589 -- TRR 211 and by the State of Hesse within the Research Cluster
ELEMENTS (Project ID 500/10.006).

\bibliographystyle{apsrev4-1}
\bibliography{Covariant_kinetic}

\end{document}